\documentclass[11pt,a4paper]{article}
\usepackage{a4wide}

\usepackage[dvipdfmx,hiresbb]{graphicx} 
\usepackage{mediabb}                    

\usepackage{amsmath,amssymb}
\usepackage{color}
\usepackage[bookmarks,bookmarksnumbered]{hyperref}
\usepackage{cellspace}
\usepackage{tikz}
\usetikzlibrary{fadings}
\usetikzlibrary{patterns}
\usetikzlibrary{shadows.blur}
\usetikzlibrary{shapes}

\usepackage{amsmath,amssymb}
\usepackage{mathtools}
\usepackage{cite}
\usepackage{authblk}
\usepackage{mathrsfs}
\usepackage{color}
\usepackage{bm}

\title{String geometry phenomenology}
\author[1]{Matsuo Sato\footnote{msato@hirosaki-u.ac.jp}}
\author[2]{Maki Takeuchi\footnote{maki\_t@yamaguchi-u.ac.jp}}
\affil[1]{\small \textit{Graduate School of Science and Technology, Hirosaki University
Bunkyo-cho 3, Hirosaki, Aomori 036-8561, Japan}}
\affil[2]{\small \textit{Graduate School of Sciences and Technology for Innovation, Yamaguchi University, Yamaguchi-shi, Yamaguchi 753-8512, Japan}}

\date{}

\begin{document}

%

{\let\newpage=\relax \maketitle}

\begin{abstract}
Recently, a potential for string backgrounds is obtained from string geometry theory, which is a candidate for the non-perturbative formulation of string theory. 
By substituting a string phenomenological model with free parameters to the potential, one obtains a potential for the free parameters, whose minimum determines the free parameters. The model with the determined parameters is the ground state in the model.  This will be the local minimum in a partial region of the model in the string theory landscape. By comparing it with the other local minimum, one can determine which model is near the minimum of the potential for string backgrounds, that will be the true vacuum in string theory, in the sense of the values of the potential. We will be able to find the true vacuum in string theory through a series of such researches. In this paper, we perform this analysis of a certain simple heterotic non-supersymmetric model explicitly, where the six-dimensional internal spaces are products of two-dimensional spaces of constant curvatures, and the generation number of massless fermions is given by the flux quantization numbers. As a result, we obtain a constraint between the compactification scale and the flux quanta.

\end{abstract}

\newpage  

\tableofcontents   

\newpage

\section{Introduction}

The higher-dimensional theories, such as string theory, are promising extensions beyond the Standard Model, with the potential to account for phenomena such as the existence of three generations of quarks and leptons \cite{Sakamoto:2020pev,Abe:2008sx,Abe:2015yva,Libanov:2000uf,Frere:2000dc,PhysRevD.65.044004,PhysRevD.73.085007,Gogberashvili:2007gg,Guo:2008ia,PhysRevLett.108.181807}, the origin of the flavor structure, including mass hierarchies \cite{Cremades:2004wa,Abe:2008sx,Arkani-Hamed:1999ylh,Dvali:2000ha,Gherghetta:2000qt,Kaplan:2000av,Huber:2000ie,Kaplan:2001ga,Fujimoto:2012wv,PhysRevD.97.115039,PhysRevD.90.105006} and sources of CP violation \cite{PhysRevD.88.115007,PhysRevD.90.105006,Kobayashi:2016qag,Buchmuller:2017vho,PhysRevD.97.075019}. 

Since the mid-1980s, supersymmetric compactifications have been the primary focus of study. However, supersymmetry has yet to be observed in collider experiments, suggesting that low-energy supersymmetry might not be a necessary requirement. Consequently, there has been a growing body of work in recent years focusing on non-supersymmetric solutions.
In \cite{Tsuyuki:2021xqu}, exact non-supersymmetric solutions to the equations of motion of heterotic supergravity are found.

One of the fundamental issues in compactification, regardless of the presence or absence of supersymmetry, is the inability to determine the free parameters.
The swampland conjecture has emerged as a topic of significant interest in this context. The Swampland conditions \cite{Vafa:2005ui} provide constraints on the parameter space of effective field theories \cite{Palti:2019,vanBeest:2021lhn,Grana:2021zvf,Agmon:2022thq,Hamada:2024gzi}. 
As shown in \cite{Hamada:2024gzi}, the swampland conjecture imposes constraints on the parameter space; however, it still remains difficult to determine the values of the parameters.

In string theory, an enormous number of perturbatively stable vacua - estimated to exceed ($>10^{500}$)
- are known to exist. This vast set of vacua is referred to as the string theory landscape. Since perturbative string theories are formulated only around local minima, they are unable to uniquely select the true vacuum. In contrast, it is believed that a non-perturbative formulation of string theory could identify the true vacuum.

A fundamental problem in string theory is to determine the six-dimensional internal space within a non-perturbative formulation. Accordingly, clarifying the nature of spacetime in string theory provides a crucial insight into the structure of a non-perturbative framework. In perturbative string theory, space-time is treated as a continuum composed of points, while particles are described as one-dimensional extended objects, namely strings. Given that quantum gravitational effects are expected to induce fluctuations in the structure of space-time, it is natural to hypothesize that, in a non-perturbative formulation, space-time itself is likewise composed of strings. This principle underlies the framework of string geometry theory \cite{Sato:2017qhj}, where both matter and spacetime emerge from fundamentally string-theoretic degrees of freedom.

String geometry theory is one of the candidates of a non-perturbative formulation of string theory. In this theory, the ``classical'' action is almost uniquely determined by T-symmetry, which is a generalization of the T-duality \cite{Sato:2023lls}, where  the parameter of ``quantum'' corrections $\beta$ in the path-integral of the theory is independent of that of quantum corrections $\hbar$ in the perturbative string theories.
We distinguish the effects of $\beta$ and $\hbar$ by putting " " like "classical" and "loops" for tree level and loop corrections with respect to $\beta$, respectively, whereas by putting nothing like classical and loops for tree level and loop corrections with respect to $\hbar$, respectively.
A non-renormalization theorem in string geometry theory states that there is no ``loop'' correction \cite{Sato:2025wfc}
. Thus, there is no problem of non-renormalizability, although the theory is defined by the path-integral over the fields including a metric on string geometry. 
No ``loop'' correction is also the reason why the complete path-integrals of the all-order perturbative strings in general string backgrounds are derived from the ``tree''-level two-point correlation functions in the perturbative vacua in \cite{Sato:2017qhj, Sato:2019cno, Sato:2020szq, Sato:2022owj, Sato:2022brv, Sato:2023lls, Nagasaki:2023fnz, Nagasaki:2025tmi, Kudo}. 
Furthermore, a non-perturbative correction in string coupling with the order $e^{-1/g_s^2}$ is given by a transition amplitude representing a tunneling process between the semi-stable vacua  in the ``classical'' potential by an ``instanton'' in the theory \cite{Sato:2025wfc}.
From this effect, a generic initial state will reach the minimum of the potential. 
Therefore, it is reasonable to conjecture that the ``classical'' potential restricted to the perturbative vacua, called the potential for string backgrounds, in string geometry theory represent the string theory landscape and the minimum of the potentials gives the true vacuum in string theory \cite{Sato:2025wfc, Nagasaki:2023fnz, Nagasaki:2025tmi, Kudo}.

Because the potential for string backgrounds represents the energies of string backgrounds, one can compare the energies of them. Thus, by substituting a model with free parameters for string backgrounds to the potential for string backgrounds, a potential for the free parameters is obtained. The minimum of the potential determines the free parameters. The model with the determined parameters is the ground state in the model. This will be the local minimum in a partial region of the model in the string theory landscape. By comparing it with the other local minimum, one can determine which model is near the minimum of the potential for string backgrounds, which will be the true vacuum in string theory, in the sense of the values of the potential. We will be able to find the true vacuum in string theory through a series of such researches. In this paper, we will perform this analysis of a certain simple heterotic non-supersymmetric model found in \cite{Tsuyuki:2021xqu,Takeuchi:2023egl} explicitly.

This paper is organized as follows. In Section \ref{sec:heterocom}, we briefly review the heterotic sector of string geometry theory and the potential for string backgrounds. In Section \ref{sec:setup}, we also briefly review the model of compactifications of heterotic supergravity. In Section \ref{sec:calculationpotential}, we analyze the potential for the parameters given by substituting the model to the potential for string backgrounds.  
By searching for the minimum of the potential for the parameters, we obtain a nontrivial relation between the compactification scale and the flux quantization number. Section \ref{sec:conclusion} is devoted to the conclusion and discussion.

\section{Brief review on the heterotic sector of string geometry theory and the potential for string backgrounds}
\label{sec:heterocom}
String manifold is constructed by patching open sets in string model space $E= \cup_{T} E_T$, where $T$ runs  IIA, IIB, SO(32) I, SO(32) het, and $E_8 \times E_8$ het.
Here, we summarize the definition of the heterotic sector of the string model space, $E_{G \mbox{het}}$ where $G$ runs $SO(32)$ and $E_8 \times E_8$. 
First, one of the coordinates of the model space is spanned by string geometry time $\bar{\tau} \in \bold{R}$ and another is spanned by heterotic super Riemann surfaces $\bar{\bold{\Sigma}} \in \mathcal{M}_{\mbox{het}}$ \cite{Witten:2012bg, Witten:2012ga,Witten:2015hwa}. 
On each  $\bar{\bold{\Sigma}}$,  
 a global time is defined canonically and uniquely by the real part of the integral of an Abelian differential \cite{Krichever:1987qk}.
We identify this global time as $\bar{\tau}$ and restrict $\bar{\bold{\Sigma}}$ to a $\bar{\tau}$ constant hyper surface, and obtain $\bar{\bold{\Sigma}}|_{\bar{\tau}}$. 
An embedding of $\bar{\bold{\Sigma}}|_{\bar{\tau}}$ to $\mathbb{R}^{d}$  is parametrized by the other coordinates 
${\boldsymbol X}_{G}^{(\mu \bar\sigma  \bar\theta)}(\bar{\tau})=X^{\mu}(\bar\sigma, \bar\tau)+ \bar{\theta} \psi^{\mu}(\bar\sigma, \bar\tau) $ where $\mu=0, 1, \cdots d-1$ and $\psi^{\mu}$ is a Majorana fermion, and $\bm X_{LG}^{(A \bar\sigma  \bar\theta^-)}(\bar\tau) =  \bar\theta^{-} \lambda_{G}^A(\bar\sigma, \bar\tau)$ where $A= 1, \cdots 32$ and 
$\bar\theta^{-}$ has the opposite chirality to $\bar\theta$. 
We abbreviate $G$ of $X^{\mu}$ and $\psi^{\mu}$. 

We can define worldsheet fermion numbers of states in a Hilbert space because the states consist of the fields over the local coordinates ${\boldsymbol X}_{G}^{(\mu \bar\sigma  \bar\theta)}(\bar{\tau})=X^{\mu}(\bar\sigma, \bar\tau)+ \bar{\theta} \psi^{\mu}(\bar\sigma, \bar\tau)$ and $\bm X_{LG}^{(A \bar\sigma  \bar\theta^-)}(\bar\tau) =  \bar\theta^{-} \lambda_{G}^A(\bar\sigma, \bar\tau)$. 
For $G=SO(32)$, we take periodicities 
\begin{equation}
\lambda_{SO(32)}^A(\bar{\tau}, \bar{\sigma}+2\pi)=\pm \lambda_{SO(32)}^A(\bar{\tau}, \bar{\sigma}) \quad (A=1, \cdots 32)
\end{equation}
with the same sign on all 32 components. 
We define the Hilbert space in these coordinates by the GSO projection of the states with $e^{\pi i F}=1$ and $e^{\pi i \tilde{F}}=1$, where $F$ and $\tilde{F}$ are the numbers of left- and right- handed fermions $\lambda^A_{SO(32)}$ and $\psi^{\mu}$, respectively. 
For $G=$ $E_8 \times E_8$, the periodicity is given by 
\begin{eqnarray}
\lambda_{E_8 \times E_8}^A(\bar{\tau}, \bar{\sigma}+2\pi)= 
\left\{
\begin{array}{c}
\eta \lambda_{E_8 \times E_8}^A(\bar{\tau}, \bar{\sigma}) \quad (1 \leqq A \leqq 16) \\
\eta' \lambda_{E_8 \times E_8}^A(\bar{\tau}, \bar{\sigma}) \quad (17 \leqq A \leqq 32), 
\end{array}
\right.
\end{eqnarray}
with the same sign $\eta(= \pm 1)$ and $\eta'(= \pm 1)$ on each 16 components.  
The GSO projection is given by $e^{\pi i F_1}=1$, $e^{\pi i F_2}=1$ and $e^{\pi i \tilde{F}}=1$, where $F_1$, $F_2$ and $\tilde{F}$ are the numbers of $\lambda_{E_8 \times E_8}^{A_1}$ ($A_1=1, \cdots, 16$), $\lambda_{E_8 \times E_8}^{A_2}$ ($A_2=17, \cdots, 32$) and $\psi^{\mu}$, respectively.

\begin{figure}[htb]
\centering
\includegraphics[width=3cm]{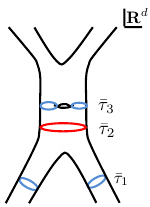}
\caption{Various string states. The red and blue lines represent one string and two strings, respectively.}
\label{states}
\end{figure}

Because the bosonic part of $\bar{\bold{\Sigma}}|_{\bar{\tau}}$ is isomorphic to $ S^1 \cup S^1 \cup \cdots \cup S^1$ and $\bold{X}_{G}(\bar{\tau}): \bar{\bold{\Sigma}}|_{\bar{\tau}} \to \mathbb{R}^{d}$, $[\bar{\bold{\Sigma}},  \bold{X}_{G}(\bar{\tau}), \bm X_{LG}(\bar\tau), \bar{\tau}]$  represent many-body strings in $\mathbb{R}^{d}$ as in Fig. \ref{states}.
Heterotic model space $E_{G \mbox{het}}$  is defined by the collection of $[\bar{\bold{\Sigma}},  \bold{X}_{G}(\bar{\tau}), \bm X_{LG}(\bar\tau), \bar{\tau}]$ by considering all the  $\bar{\bm\Sigma}$, all the values of $\bar{\tau}$, and all the $\bold{X}_{G}(\bar{\tau})$ and $\bm X_{LG}(\bar\tau)$: 
$E_{G \mbox{het}} = \{[\bar{\bold{\Sigma}},  \bold{X}_{G}(\bar{\tau}), \bm X_{LG}(\bar\tau), \bar{\tau}]\}$.
In the following,  instead of $[\bar{{\boldsymbol \Sigma}}, \bold{X}_{G}(\bar{\tau}), \bm X_{LG}(\bar\tau), \bar{\tau}]$,  we denote $[\bar{\bold{E}}, \bold{X}_{G}(\bar{\tau}), \bm X_{LG}(\bar\tau), \bar{\tau}]$, 
where $\bar{\bold{E}}$ is the worldsheet super vierbein on $\bar{{\boldsymbol \Sigma}}$ \cite{Brooks:1986uh}, 
because giving a super Riemann surface is equivalent to giving a super vierbein up to super diffeomorphism and super Weyl transformations.

We consider all manifolds obtained by gluing open subsets of the model space $E_{G \mbox{het}}$ with the coordinate $\bar{\bold{E}}$ fixed as a global coordinate, and we call them heterotic string manifolds. In other words, when gluing together open subsets of the model space $E_{G \mbox{het}}$, the moduli space of the Riemannian surfaces $\{\bar{\bold{E}}\}=\mathcal{M}_{\mbox{het}}$ are taken to be common. The path integral of the theory will later be defined on string manifolds including these heterotic string manifolds.

The summations over $(\bar\sigma, \bar\theta)$ and $(\bar\sigma,\bar\theta^{-})$ are defined by 
$\displaystyle\int d\bar\sigma d\bar\theta\hat{\bm E}(\bar\sigma, \bar\tau, \bar\theta)$ and 
$\displaystyle\int d\bar\sigma d\bar\theta^{-}\bar e(\bar\sigma, \bar\tau)$, respectively. $\hat{\bm E}(\bar{\sigma}, \bar{\tau}, \bar{\theta})\coloneqq 
(1/\bar{n})\bar{\bm E}(\bar{\sigma}, \bar{\tau}, \bar{\theta})$, where $\bar n$ is the lapse function of the two-dimensional metric.
This summation is transformed as a scalar under $\bar\tau \mapsto \bar\tau'(\bar\tau, \bm X_{G}(\bar\tau), \bm X_{L G}(\bar\tau))$ and invariant under a supersymmetry transformation $(\bar{\sigma}, \bar{\theta}) \mapsto (\bar{\sigma}'(\bar{\sigma}, \bar{\theta}), \bar{\theta}'(\bar{\sigma}, \bar{\theta}))$. 
As a result, the heterotic part of any action is invariant under this $\mathcal N = (1,0)$ supersymmetry transformation because all the indices are contracted by the summations.

  $\bar{\bold{E}}$ is a global coordinate and does not transform under general coordinate transformations; therefore, fields do not carry indices along $\bar{\bold{E}}$. We use the Einstein notation for the index $I = \{d, (\mu \bar\sigma  \bar\theta), (A \bar\sigma  \bar\theta^-) \}$ with respect to the other coordinates $\bar\tau$, $\boldsymbol X_{G}^{(\mu \bar\sigma  \bar\theta)}(\bar{\tau})$ and  $\bm X_{LG}^{(A \bar\sigma  \bar\theta^-)}(\bar\tau)$. 
Thus, 
an explicit form of the line element is given  in the same way as in the finite dimensional case by
\begin{align}\label{LineElement}
&ds^2(\bar{\bm E}, \bm X_{G}(\bar\tau), \bm X_{LG}(\bar\tau), \bar\tau)\nonumber\\
&= G(\bar{\bm E}, \bm X_{G}(\bar\tau),  \bm X_{LG}(\bar\tau), \bar\tau)_{IJ} 
d \bm X_{G}^I(\bar\tau)
d \bm X_{G}^J(\bar\tau)
\nonumber\\ 
&= G(\bar{\bm E}, \bm X_{G}(\bar\tau),  \bm X_{LG}(\bar\tau), \bar\tau)_{dd} (d\bar\tau)^2\nonumber\\ 
&\quad
+ 2d\bar\tau\int d\bar\sigma d\bar\theta\hat{\bm E}
\sum_\mu G(\bar{\bm E}, \bm X_{G}(\bar\tau), \bm X_{LG}(\bar\tau), \bar\tau)_{d(\mu \bar\sigma\bar\theta)} d\bm X_{G}^{\mu}(\bar{\sigma}, \bar{\tau}, \bar{\theta}) \nonumber \\ 
&\quad
 +2 d\bar\tau\int d\bar\sigma d\bar\theta^{-} \bar e\sum_AG(\bar{\bm E}, \bm X_{G}(\bar\tau), \bm X_{LG}(\bar\tau,\bar\tau)_{d(A\bar\sigma\bar\theta^{-})} d \bm X_{LG}^A(\bar\sigma, \bar\tau, \bar\theta^{-})\nonumber\\
&\quad
 +\int d\bar{\sigma} d\bar{\theta} \hat{\bm E}  \int d\bar\sigma'd\bar\theta'\hat{\bm E}'   
 \sum_{\mu, \mu'} G(\bar{\bm E},\bm X_{\hat D_{T}}(\bar\tau), \bm X_{LG}(\bar\tau), \bar{\tau})_{ \; (\mu \bar{\sigma} \bar{\theta})  \; (\mu' \bar{\sigma}' \bar{\theta}')} d\bm X_{G}^{\mu}(\bar{\sigma}, \bar{\tau}, \bar{\theta})d\bm X_{G}^{\mu'}(\bar\sigma',\bar\tau, \bar\theta')\nonumber\\
&\quad
 +\int d\bar\sigma d\bar{\theta}\hat{\bm E}\int d\bar\sigma' d\bar\theta^{-}\bar e'
  \sum_{\mu, A} G(\bar{\bm E}, \bm X_{G}(\bar\tau),\bm X_{LG}(\bar\tau), \bar\tau)_{(\mu \bar\sigma\bar\theta)(A\bar\sigma'\bar\theta^{-})}d\bm X_{G}^\mu(\bar\sigma,\bar\tau,\bar\theta) 
 d\bm X_{LG}^A(\bar\sigma',\bar\tau,\bar\theta^{-})\nonumber\\
&\quad
 +\int d\bar{\sigma} d \bar\theta^{-}\bar{e}  \int d\bar{\sigma}'  d \bar{\theta}^{'-} \bar{e}'  \sum_{A, A'} G(\bar{\bm E},\bm X_{G}(\bar\tau),\bm X_{LG}(\bar\tau),\bar\tau)_{(A \bar\sigma\bar\theta^{-})(A'\bar\sigma'\bar\theta'^{-})}
d\bm X_{LG}^A(\bar\sigma,\bar\tau,\bar\theta^{-})d\bm X_{LG}^{A'}(\bar\sigma', \bar\tau, \bar\theta'^{-}). 
\end{align}
Here, we should note that the fields are functionals of $\bar{\bm E}$. 
The inverse metric $\bm G^{IJ}(\bar{\bm E}, \bm X_{G}(\bar\tau), \bm X_{LG}(\bar\tau), \bar\tau)$ is given by 
$\bm G_{IJ}\bm G^{JK} = \bm G^{KJ}\bm G_{JI}= \delta^{K}_{I}$, where 
$\delta^{d}_{d}=1$,  
$\delta_{\mu \bar\sigma \bar\theta}^{\mu' \bar\sigma' \bar\theta'} 
 = \frac1{\hat{\bm E}}\delta_\mu^{\mu'}\delta(\bar\sigma - \bar\sigma') \delta(\bar\theta - \bar\theta')$, 
$\delta_{A \bar{\sigma} \bar{\theta}^-}^{A' \bar{\sigma}' \bar{\theta}^{'-}} 
 = \frac1{\bar e}\delta_{A}^{A'}\delta(\bar\sigma - \bar\sigma') \delta(\bar\theta^{-}-\bar\theta'^{-})$. 
The dimensions of string manifolds, which are infinite dimensional manifolds, are formally given by the trace of ``1'', $\delta^M_M = D+1$, where 
$\displaystyle D\coloneqq \int d\bar{\sigma} d\bar{\theta} \hat{\bm E} \delta^{(\mu\bar\sigma \bar\theta)}_{(\mu\bar\sigma\bar\theta)}+\int d\bar\sigma d\bar\theta^{-}\bar e \delta_{A \bar{\sigma} \bar{\theta}^-}^{A \bar{\sigma} \bar{\theta}^{-}}$. Thus, we treat $D$ as regularization parameter and will take $D \to\infty$ later. The scalar $\varPhi(\bar h, X(\bar\tau), \bar\tau)$
and tensors $\mathcal B_{IJ}(\bar h, X(\bar\tau), \bar\tau), \cdots$ are also defined in the same way as in the finite dimensional case.

String geometry theory is defined by a partition function
\begin{align}
Z = \int \mathcal D\bm G\mathcal D\bm\Phi  \mathcal{D}\bm B\mathcal D\bm A\mathcal D\bm{\mathcal C}e^{i S},
\label{pathint}
\end{align}
where the action is given by
\begin{equation}
S = \int\mathcal D\bar\tau \mathcal D\bm E\mathcal D\bm X_{T}\sqrt{-\bm{\mathcal G}} 
 \bigg(e^{-2\bm\varPhi} \Big(-\bm{\mathcal R} 
 - 4\bm\nabla_I\bm\varPhi\bm\nabla^I\bm\varPhi 
 + \frac12\vert\tilde{\bm{\mathcal H}}\vert^{2} +\frac{\alpha'}{4} \mathrm{tr}(\vert\bm{\mathcal F}\vert^2)\Big)
 + \frac12 \sum_{p=1}^\infty \vert\tilde{\bm{\mathcal F}}_p\vert^2     
\bigg), 
\label{action of bos string-geometric model}
\end{equation}
where
$|\tilde{\bm{\mathcal H}}|^2 =
 (1/3!)\bm{\mathcal G}^{I_1J_1}\bm{\mathcal G}^{I_2J_2}\bm{\mathcal G}^{I_3J_3} 
 \tilde{\bm{\mathcal H}}_{I_1I_2I_3} \tilde{\bm{\mathcal H}}_{J_1J_2J_3}$ for example.
 The action contains no derivatives with respect to $\bar{\bold{E}}$, 
because the fields do not carry indices with respect to $\bar{\bold{E}}$ and the action is generally covariant.
Thus, the theory is non-dynamical along $\bar{\bold{E}}$.
The equations of motion of this action \eqref{action of bos string-geometric model} can be consistently truncated to the ones of all the ten-dimensional  supergravities, namely type IIA, IIB, SO(32) type I, and type  SO(32) and $E_8 \times E_8$ heterotic supergravities \cite{Honda:2020sbl, Honda:2021rcd}. That is, this model includes all the superstring backgrounds.  Moreover, the action \eqref{action of bos string-geometric model} is strongly constrained by T-symmetry in string geometry theory, which is a generalization of T-duality among perturbative vacua in superstring theory \cite{Sato:2023lls}.
The action consists of fields on Riemannian string manifolds: a scalar curvature  $\bm{\mathcal R}$ of a metric $\bm{\mathcal G}_{I_1I_2}$, a scalar field $\bm\varPhi$, $p$-forms $\tilde{\bm{\mathcal F}}_p = \bm{\mathcal F}_p+\bm{\mathcal H}_3\wedge\bm{\mathcal C}_{p-3}$ where  $\bm{\mathcal F}_p$ are field strengths of ($p-1$)-form fields $\bm{\mathcal C}_{p-1}$,  $\tilde{\bm{\mathcal H}} = d \bm{\mathcal B} -\bm{\mathcal \omega}_3$ where $d \bm{\mathcal B}$ is a field strength of  a two-form field $\bm{\mathcal B}$, 
$\bm{\mathcal \omega}_3 = \mathrm{tr}(\bm {\mathcal A}\wedge d\bm{\mathcal  A} 
 - (2i/3)\bm{\mathcal  A} \wedge\bm{\mathcal  A} \wedge\bm{\mathcal  A})$, and 
$\bm{\mathcal  A}$ is an $N \times N$ Hermitian gauge field, whose field strength is given by $\bm{\mathcal  F}$. It is natural that the backgrounds of perturbative string theory are included in the expectation values of the fields in a non-perturbative formulation of string theory.  Actually, the fundamental fields in string geometry theory are extensions of those in the ten-dimensional supergravities. In order to minimize the number of the fundamental fields, the theory does not include such extensions of the massive modes in string theory.   However, the massive modes are included non-trivially in the theory because the perturbative string theory is derived from string geometry theory as in \cite{Sato:2017qhj, Sato:2019cno, Sato:2020szq, Sato:2022owj, Sato:2022brv, Nagasaki:2023fnz}.

In the following, we fix $\text{T} = SO(32) \text{ hetero or }E_8 \times E_8 \text{ hetero}$, namely we choose heterotic chatrs, where $\bm C_p=0$.
In these charts, the action \eqref{action of bos string-geometric model} becomes 
\begin{equation}\label{eq:sec3_adction}
S = \int\mathcal D\bar\tau \mathcal D\bm E\mathcal D\bm X_{G}\mathcal D\bm X_{LG}
\sqrt{-\bm{\mathcal G}}e^{-2\bm\varPhi}
\Big(-{\bm{\mathcal R}} - 4\bm\nabla_I{\bm\varPhi}\bm\nabla^I{\bm\varPhi} 
 + \frac12 |\tilde{\bm{\mathcal H}}|^2 + \frac{\alpha'}{4}\mathrm{tr}|\bm{\mathcal F}|^2\Big),
\end{equation}
where
$I = \{d,(\mu\bar\sigma\bar\theta),(A\bar\sigma\bar\theta^{-})\}$.

 Here, we consider fluctuations around "classical" backgrounds in the action of string geometry theory. If we fix certain "classical" backgrounds including heterotic string backgrounds in the ten dimensions, $g_{\mu\nu}(x)$, $B_{\mu \nu}(x)$. $\Phi(x)$ and $A_{\mu}(x)$ as parameters,
 \begin{subequations}\label{eq:sec3_backgrounds}
\begin{align}
\bm{\mathcal G}_{IJ} &=\bar{\bm G}_{IJ}, 
\label{eq:sec3_backgrounds_G}\\
\bm{\mathcal B}_{IJ}& =\bar{\bm B}_{IJ} 
\label{eq:sec3_backgrounds_B}\\
\bm\varPhi &= \bar{\bm\varPhi}
\label{eq:sec3_backgrounds_Phi}\\
\bm{\mathcal A}_{I}& = \bar{\bm A}_{I},
\label{eq:sec3_backgrounds_C}
\end{align}
\end{subequations}
where
\begin{subequations}\label{eq:sec3_condition2}
\begin{align}
\bar{\bf G}_{dd}&= {\bf G}_{dd} = e^{2\phi[G,B,\Phi, A; X]}  \\
\bar{\bf G}_{d(\mu\bar{\sigma}\bar{\theta})}&=0 \\
\bar{\bf G}_{d(A\bar{\sigma}\bar{\theta}^-)}&=0 \\
\bar{\bf G}_{(\mu\bar{\sigma}\bar{\theta})(\mu'\bar{\sigma}'\bar{\theta}')}&=
{\bf G}_{(\mu\bar{\sigma}\bar{\theta})(\mu'\bar{\sigma}'\bar{\theta}')}
=
\frac{\bar{e}^3}{\sqrt{\bar{h}}}\,
G_{\mu\nu}({\bf X}_{G}(\bar{\sigma}, \bar{\theta}))
\delta_{\bar{\sigma}\bar{\sigma}'}\,\delta_{\bar{\theta}\bar{\theta}'} \\
\bar{\bf G}_{(\mu\bar{\sigma}\bar{\theta})(A\bar{\sigma}'\bar{\theta'}^-)}&=0\\
\bar{\bf G}_{(A\bar{\sigma}\bar{\theta}^-)(A'\bar{\sigma}'\bar{\theta'}^-)}&=
\bf G_{(A\bar{\sigma}\bar{\theta}^-)(A'\bar{\sigma}'\bar{\theta'}^-)}
=\frac{\bar{e}^3}{\sqrt{\bar{h}}}\,\delta_{AA'}\,
\delta_{\bar{\sigma}\bar{\sigma}'}\,\delta_{\bar{\theta}^-\bar{\theta'}^-}
\\
\bar{\bf B}_{d(\mu\bar{\sigma}\bar{\theta})}&=0  \\
\bar{\bf B}_{d(A\bar{\sigma}\bar{\theta}^-)}&=0 \\
\bar{\bf B}_{(\mu\bar{\sigma}\bar{\theta})(\mu'\bar{\sigma}'\bar{\theta}')}&=
{\bf B}_{(\mu\bar{\sigma}\bar{\theta})(\mu'\bar{\sigma}'\bar{\theta}')}
=\frac{\bar{e}^3}{\sqrt{\bar{h}}}\,
B_{\mu\nu}({\bf X}_{G}(\bar{\sigma}, \bar{\theta}))
\delta_{\bar{\sigma}\bar{\sigma}'}\,\delta_{\bar{\theta}\bar{\theta}'} \\
\bar{\bf B}_{(\mu\bar{\sigma}\bar{\theta})(A\bar{\sigma}'\bar{\theta'}^-)}&=0\\\bar{\bf B}_{(A\bar{\sigma}\bar{\theta}^-)(A'\bar{\sigma}'\bar{\theta'}^-)}&=0
\\
\bar{\varPhi}&=\varPhi= \int d \bar{\sigma}\,d\bar{\theta}\,\hat{\bf E}\,\Phi({\bf X}_{G}(\bar{\sigma}, \bar{\theta})),\\
\bar{\bf A}_{d}&=0 \\
\bar{\bf A}_{(\mu\bar{\sigma}\bar{\theta})}&=
{\bf A}_{(\mu\bar{\sigma}\bar{\theta})}
=
\frac{\bar{e}^3}{\sqrt{\bar{h}}}\,
A_{\mu}({\bf X}_{G}(\bar{\sigma}, \bar{\theta}))\\
\bar{\bf A}_{(A\bar{\sigma}\bar{\theta}^-)}&=0, 
\end{align}
 \end{subequations}
 two-point correlation functions of the scalar modes of the fluctuations of the metrics give path-integrals of the perturbative heterotic strings on the string backgrounds up to any order of the string coupling. Thus, the "classical" backgrounds are identified with the perturbative heterotic vacua in string theory. Moreover, the string backgrounds are restricted to the solutions to the equations of motion of the ten-dimensional heterotic supergravity by the consistency of the fluctuations, namely the Weyl invariance of the perturbative strings. The same is true in the case of type I and type II as in \cite{Kudo}. 

 As discussed in Introduction, it is reasonable to conjecture that the ``classical'' potential restricted to the perturbative vacua, called the potential for string backgrounds, in string geometry theory represent the string theory landscape and the minimum of the potentials gives the true vacuum in string theory \cite{Sato:2025wfc, Nagasaki:2023fnz, Nagasaki:2025tmi, Kudo}.
The potential for string backgrounds is given by the minus of the "classical" action restricted to the perturbative vacua, because the perturbative vacua are independent of the time in string geometry theory as in (\ref{eq:sec3_condition2}). Among solutions to the equations of motion of supergravities, the minimum of the potentials will be the true vacuum in string theory.

 Explicitly, the potential for heterotic string backgrounds in the Einstein frame with the warped compactification in a particle limit $\bm X_{G}^{(\mu\bar\sigma\bar\theta)}  \to x^{\mu}$, is given by 
\begin{align}
V_\text{warp} 
&= \int d^6y\sqrt{g}\Big[
 e^\phi\Big(R - \frac12e^{-\Phi+4\rho}|\tilde H|^2 -\frac{\alpha'}{4} e^{-\frac12\Phi+2\rho}\mathrm{tr}|F|^2
 + 2\nabla^2\rho - 8(\partial \rho)^2 + \frac12\nabla^2\Phi - \frac12(\partial\Phi)^2 \nonumber \\
& \qquad \qquad \qquad \qquad -2\nabla^2\phi - 2(\partial\phi)^2 - 4\partial_\mu\Phi\partial^\mu\phi\Big)  -  fe^{\Phi+\frac12\phi}(\nabla^2\phi + (\partial\phi)^2 )\nonumber \\
& \qquad \qquad \qquad 
+P\Big(\nabla^2\phi + \frac{9}{2}(\partial\phi)^2
 + 7\nabla^2\Phi -3(\partial\Phi)^2 \nonumber \\
& \qquad \qquad \qquad  \qquad \quad +2R -e^{-\Phi+4\rho}|\tilde H|^2-\frac{\alpha'}{2} e^{-\frac12\Phi+2\rho}\mathrm{tr}|F|^2  +4\nabla^2\rho - 16(\partial \rho)^2  \Big) \nonumber\\
& \qquad \qquad \qquad 
+ Q \Big(\nabla^2 f + 2\partial_m\Phi\partial^m f 
 - e^{-\Phi+\frac12\phi}(\nabla^2\phi + (\partial\phi)^2 ) \Big) \Big] .
\label{eq:sec5_diffeq_wcomp_f_1}
\end{align}
where $P$ and $Q$ are the Lagrange multipliers that impose the conditions determining $\phi$ and $f$, respectively \cite{Nagasaki:2023fnz, Nagasaki:2025tmi, Kudo}. In the warped compactification,  
the metric is given by 
\begin{equation}\label{eq:sec4_metric}
ds^2 = e^{2\rho(y)}\eta_{\alpha \beta}dx^{\alpha}dx^{\beta} + e^{-2\rho(y)}g_{mn}(y)dy^mdy^n,
\end{equation}
where $\alpha, \beta= 0, \cdots, 3$ and $m,n= 4, \cdots, 9$,
and the other non-zero backgrounds are given by  $B_{mn}(y)$,  $\Phi(y)$ and $A_m(y)$.

There are two ways to treat the conditions determining $\phi$ and $f$, during analysis of this potential. One way is to treat $P$ and $Q$ as independent fields and to search for the minimum. Another is to solve the conditions completely and then obtain exact or series solutions. Here, we must not approximate  fields,  for example up to the second order, when we solve the conditions, because the absolute minimum of the potential is defined globally. 

In case that models have only finite number of free parameters as in this paper, by solving the conditions, infinite degrees of freedom $P$ and $Q$ disappear, and then the problem reduces to easier problem to find the minimum among the finite degrees of freedom.

\section{Brief review on a certain simple heterotic non-supersymmetric model}
\label{sec:setup}
We review the compactification presented in Ref. \cite{Tsuyuki:2021xqu,Takeuchi:2023egl}. Our focus is on the bosonic sector of the heterotic supergravity Lagrangian \cite{Bergshoeff:1989de,Lechtenfeld:2010dr}:

\begin{align}
L=\sqrt{-g}e^{-2\Phi}\left[R+4(\nabla \Phi)^2-\frac{1}{12}H_{\mu \nu \rho }H^{\mu \nu \rho}+\frac{\alpha'}{8}R_{\mu \nu \rho \sigma}R^{\mu \nu \rho \sigma}-\frac{\alpha'}{8}\textrm{tr}(F_{\mu \nu}F^{\mu \nu })\right].
\end{align}
Here, $g$ is the determinant of the metric $g_{\mu \nu }\,(\mu, \nu  =0,\ldots,9)$. $R_{\mu \nu \rho \sigma }$ is the Riemann tensor, the Ricci scalar $R$ and Ricci tensor $R_{\mu \nu}$ are defined as  $R= g^{\nu \sigma }R_{\nu \sigma} = g^{\mu \rho }g^{\nu \sigma} R_{\mu \nu \rho \sigma}$, and $F_{\mu \nu}$ is the gauge field strength of the gauge group $E_8\times E_8$.
We take the dilaton $\Phi$ as a constant and assume that $H$ is zero,
\begin{align}
\partial_{\mu} \Phi=0,\quad H_{\mu \nu \rho}=0. \label{edp}
\end{align}
The equations of motion \cite{Becker:2009df,Lechtenfeld:2010dr} become
\begin{align}
R+\frac{\alpha'}{8}R_{\mu \nu \rho \sigma}R^{\mu \nu \rho \sigma}-\frac{\alpha'}{8}\textrm{tr}(F_{\mu \nu}F^{\mu \nu}) &=0,\label{eral}\\
R_{\mu \nu}+\frac{\alpha'}{4}R_{\mu  \rho \sigma \lambda}{R_{\nu}}^{\rho \sigma \lambda} -\frac{\alpha'}{4}\textrm{tr}(F_{\mu \rho}{F_{\nu}}^{\rho}) &=0, \label{ermn}\\
\nabla_{\mu} F^{\mu \nu}+[A_{\mu},F^{\mu \nu}] &=0. \label{efm}
\end{align}
Then, we get
\begin{align}
    R=0.
\end{align}
From the Green-Schwarz anomaly cancellation condition \cite{Green:1984sg}, it is additionally required to satisfy the following relation
\begin{align}
    0=dH=\frac{\alpha'}{4}(\textrm{tr} \mathcal{R}\wedge \mathcal{R} - \textrm{tr} F\wedge F). \label{edh}
\end{align}
We consider that the ten-dimensional spacetime is a product of following four manifolds:
\begin{align}
M^{10}=M_0\times M_1\times M_2\times M_3, \label{em10}
\end{align}
where $M_0$ is the four-dimensional Minkowski spacetime and $M_i\ (i=1,2,3)$ are two-dimensional spaces of constant curvature. 
The Riemann tensor components are
\begin{align}
R_{mnpq}^{(i)} &=\lambda_i (g_{mp}^{(i)}g_{nq}^{(i)}-g_{mq}^{(i)}g_{np}^{(i)}), \label{erm}
\end{align}
with $m,n,p,q$ being indices along the tangent space of $M_i$, and $\lambda_i$ is a constant sectional curvature. When $\lambda_i$ takes the values $\lambda_i>0$, $\lambda_i=0$, and $\lambda_i<0$, the corresponding manifolds are a sphere, a torus, and a compact hyperbolic manifold, respectively.

In this model, not only $H_3=0$ but also $\tilde{H}_3=0$. The reason is the following. One spin connection $\omega_1$ and its derivative $d\omega_1$ have the indices only from one two-dimensional internal submanifold (\ref{em10}) because of the ansatz of the metrics (\ref{erm}). Because the indices of 
$\omega_{3L}=\mbox{tr}(\omega_1 \wedge d\omega_1 -\frac{2i}{3}\omega_1 \wedge \omega_1 \wedge \omega_1)$
are completely anti-symmetric, at least two indices of it are from different two-dimensional internal submanifolds. Because the product of the matrices from the different  two-dimensional internal submanifolds is zero, $\omega_{3L}=0$.

We assume the Freund-Rubin configuration \cite{Freund:1980xh}
of the gauge field strength $F_{\mu \nu}=F_{A \mu \nu}T_A$,
\begin{align}
F_{Ai,mn} &=\sqrt{g_i}f_{Ai} \epsilon_{mn}^{(i)}, \label{efmn}
\end{align}
where $g_i=\textrm{det}(g_{mn}^{(i)})$, $f_{Ai}$ is a constant, and $\epsilon_{mn}^{(i)}$ is a Levi-Civita symbol for $M_i$. 

By the Gauss-Bonnet theorem, we get 
\begin{align} 
\int_{M_i}\frac{R^{(i)}}{2}\sqrt{g_i}d^2x=\textrm{vol}(M_i)\lambda_i=2\pi\chi_i, \label{erv}
\end{align}
where $R^{(i)}$ is the Ricci scalar of $M_i$ and $\chi_i$ is its Euler characteristic.
On the other hand, the gauge field strength satisfies the flux quantization condition \cite{Witten:1984dg}
\begin{align} 
\int_{M_i} F_{Ai} = \textrm{vol}(M_i) f_{Ai} = 2\pi n_{Ai}, \label{efv}
\end{align}
where $n_{Ai}$ is an integer. Therefore, we get
\begin{align}
f_{Ai}=\frac{n_{Ai}}{\chi_i}\lambda_i. \label{efia}
\end{align}
The equations of motion and the anomaly cancellation condition become the following integer equations:
\begin{align}
&\lambda_1+\lambda_2+\lambda_3=0, 
\label{eq1}\\
&\sum_{A=1}^{11}n_{Ai}^2 = \chi_i^2\left(1+\frac{1}{\lambda_i}\right), \label{elic}\\
&\sum_{A=1}^{11} n_{Ai}n_{Aj} = 0 \quad (i\neq j). \label{eq:bi}
\end{align}
As discussed in Ref. \cite{Tsuyuki:2021xqu,Takeuchi:2023egl}, due to these constraints, the possible compactifications are limited to $S^2\times H^2/\Gamma\times H^2/\Gamma$ or $S^2\times S^2\times H^2/\Gamma$.

The decomposition of $E_8$ and a gaugino in 248 representation for SO(10) GUT \cite{Green:2012pqa} is
\begin{align}
E_8 &\supset SO(10)\times SU(4),\\
248 &= (45,1)+(1,15)+(10,6)+(16,4)+(\overline{16},\overline{4}).
\end{align}
One generation of the standard model fermions can be in a 16 representation. The number of generations is given by the index theorem \cite{Bars:1985nz}
\begin{align}
N_{\text{gen}} &= \frac{1}{(2\pi)^3}\frac{1}{6}\left|\int \text{tr}(F^3)\right|.
\end{align} 
We take the basis of the Cartan subalgebra of $\mathfrak{su}(4)$ as
\begin{align}
\begin{split}
T_1&=
\text{diag}[-1,0,0,1],\\
T_2&=
\text{diag}[0,-1,0,1],\\
T_3&=
\text{diag}[0,0,-1,1].    
\end{split}
\end{align}
We consider that $U(1)_A\ (A=1,2,3)$ fluxes are in the $SU(4)$. The other $U(1)_A\ (A=4,\dots,11)$  fluxes in another $E_8$ are not relevant for $N_{\text{gen}}$.
We can describe the number of generations using the flux quantization number as
\begin{align}
N_{\text{gen}}=&\left|\prod_{i=1}^3\sum_{A=1}^3 n_{Ai} -\sum_{A=1}^3 \prod_{i=1}^3n_{Ai}\right|. \label{engen}
\end{align}
As shown in \cite{Takeuchi:2023egl}, in the case that the Euler characteristics of the three internal submanifolds are $(\chi_1,\chi_2,\chi_3)=(2,-2,-2)$, the number of generations is constrained to be three or less due to the conditions Eqs.\eqref{eq1}-\eqref{eq:bi}.
The important point is that the conditions \eqref{eq1}-\eqref{eq:bi} impose constraints on the values of the flux, which in turn restrict the number of generations. However, as the number of genus increases, the allowed values of the flux also become larger, and thus the constraint on the number of generations becomes weaker.
To the best of our knowledge, there has been no theoretical constraint on the genus. As a result, it has not been possible to determine which solution satisfying conditions Eqs.\eqref{eq1}-\eqref{eq:bi} should be selected.
In this paper, we treat the flux values and the constant dilaton as free parameters as  the genus depend on the flux, and will explore configurations that minimize the potential for string backgrounds in string geometry theory.

\section{Evaluation of the potential}
\label{sec:calculationpotential}

In this section, we will substitute the models introduced in the previous section into the potential for string backgrounds described in Section \ref{sec:heterocom} and evaluate the resulting expressions.
By imposing the conditions
\begin{align}
    \partial_{\mu}\Phi=0,\,R=0,\,\tilde{H}=0,\,\rho=0,
\end{align}
from the models described in the previous section into Eq.\eqref{eq:sec5_diffeq_wcomp_f_1}, we obtain the potential for our models as
\begin{align}
    V=\int d^6 x \sqrt{g} \left(-\frac{1}{2}e^{\phi}b-(2e^{\phi}+f e^{\Phi+\frac{1}{2}\phi})(\nabla^2 \phi+(\partial \phi)^2)\right),
    \label{makipotential}
\end{align}
$\phi$ and $f$ are solutions satisfying the following equations:
\begin{align}
    &\nabla^2 \phi +\frac{9}{2} (\partial \phi)^2 -b=0,\label{phieq}\\
    &\nabla^2 f -e^{-\Phi+\frac{1}{2}\phi}(\nabla^2 \phi +(\partial \phi)^2)=0,
    \label{feq}
\end{align}
where
\begin{equation}
b=\frac{\alpha^{\prime}}{2}e^{-\frac{1}{2}\Phi}{\rm{tr}}|F|^2. \label{combination}
\end{equation}
From Eq.\eqref{phieq}, the potential is rewritten as
\begin{align}
    V=\int d^6 x \sqrt{g} \left(-\frac{1}{2}e^{\phi}b-(2e^{\phi}+f e^{\Phi+\frac{1}{2}\phi})\left(\frac{7}{9}\nabla^2 \phi+\frac{2}{9}b\right)\right).
    \label{makipotential2}
\end{align}

\subsection{Solving the condition for $\phi$}
In this subsection, we will solve the condition Eq.\eqref{phieq} completely and obtain a series solution,
in order to avoid infinite degrees of freedom, as stated in the end of Section \ref{sec:heterocom}. We solve Eq.\eqref{phieq} in terms of a formal power series
$\phi=\sum_{i=0}^{\infty}\phi_i$. By substituting, we get
\begin{align}
  \sum_{i=0}^{\infty}\nabla^2\phi_i= \sum_{i=1}^{\infty} \left(-\frac{9}{2}\sum_{k+l+1=i}\partial \phi_k \partial \phi_l \right) +b .
\end{align}
Thus, the conditions to be satisfied at each order are given by
\begin{align}
    \nabla^2\phi_0&=b, \\
    \nabla^2\phi_i&=-\frac{9}{2}\sum_{k+l+1=i}\partial \phi_k \partial \phi_l \quad (i\geq 1).
\end{align}
Here, we introduce the Green function $G(x,y)$ satisfied $\nabla_x^2 G(x,y)=\delta^6(x-y)$. We obtain
\begin{align}
    \phi_0&=b\int d^6y  \sqrt{g}\,G(x,y), \label{phi0}\\
    \phi_i&=-\frac{9}{2} \int d^6y  \sqrt{g} \,G(x,y) \sum_{k+l+1=i}\partial \phi_k (y)\partial \phi_l(y).
\end{align}
We further calculate as 
\begin{align}
    \phi_i&=-\frac{9}{2} \int d^6y  \sqrt{g}\,G(x,y) \sum_{k+l+1=i}\partial \phi_k (y)\partial \phi_l(y) \notag \\
    &=\frac{9}{2} \int d^6y  \sqrt{g}\,\partial_y G(x,y) \sum_{k+l+1=i} \phi_k (y)\partial \phi_l(y)
    +\frac{9}{2} \int d^6y  \sqrt{g} \, G(x,y) \sum_{k+l+1=i} \phi_k (y)\nabla^2 \phi_l(y)
    \notag \\
    &=-\frac{9}{2} \int d^6y  \sqrt{g}\,\nabla_y^2 G(x,y) \sum_{k+l+1=i} \phi_k (y) \phi_l(y)
   - \frac{9}{2} \int d^6y  \sqrt{g} \,\partial_y G(x,y) \sum_{k+l+1=i} \partial \phi_k (y) \phi_l(y) \notag \\
    &\qquad +\frac{9}{2} \int d^6y  \sqrt{g} \, G(x,y) \sum_{k+l+1=i} \phi_k (y)\nabla^2 \phi_l(y)
    \notag \\
    &=-\frac{9}{4} \int d^6y   \sqrt{g}\,\nabla_y^2 G(x,y) \sum_{k+l+1=i} \phi_k (y) \phi_l(y)
    +\frac{9}{2} \int d^6y  \sqrt{g}\, G(x,y) \sum_{k+l+1=i} \phi_k (y)\nabla^2 \phi_l(y) \notag \\
    &=-\frac{9}{4} \sum_{k+l+1=i} \phi_k  \phi_l +\frac{9}{2} \int d^6y  \sqrt{g}\, G(x,y) \sum_{k+l+1=i} \phi_k (y)\nabla^2 \phi_l(y) .
    \label{phi_i}
\end{align}
In the last equality, we assume the relation $\nabla_y^2 G(x,y)=\delta^6(x-y)$.
The explicit solutions for $\phi_i$ are given by
\begin{align}
    \phi_0&=b\Lambda_1(x), \\
    \phi_1&=-\frac{9}{2}b^2\left(\frac{1}{2}\Lambda_1^2(x)-\Lambda_2(x)\right), \\
    \phi_2&=-\left(-\frac{9}{2}\right)^2 b^3\left(-\Lambda_3(x)+\Lambda_2(x)\Lambda_1(x)-\frac{1}{3}\Lambda_1^3(x)\right), \\
    \phi_3&=\left(-\frac{9}{2}\right)^3 b^4\left(-\Lambda_4(x)+\Lambda_3(x)\Lambda_1(x)+\frac{1}{2}\Lambda_2^2(x)-\Lambda_2(x)\Lambda_1(^2x)+\frac{1}{4}\Lambda_1^4(x)\right), \\
    &\cdots \notag \\
     \phi_n&=\left(-\frac{9}{2}\right)^n (-b)^{n+1} \sum_{i_1+i_2+\cdots+i_k=n+1}\frac{(-1)^k}{k}
     \Lambda_{i_1}\Lambda_{i_2}\cdots \Lambda_{i_k},
     \label{phi_n}
\end{align}
where $\Lambda_j$ is defined by $\Lambda_1(x)=\int d^6y  \sqrt{g} \,G(x,y)$, $\Lambda_2(x)=\int d^6y  \sqrt{g} \int d^6z  \sqrt{g}\,G(x,y)G(y,z),\cdots,\Lambda_j(x)=\int d^6y_1  \sqrt{g}\cdots \int d^6y_j  \sqrt{g}\,G(x,y_1)G(y_1,y_2)\cdots G(y_{j-1},y_j)$. Details of the calculation are given in the Appendix \ref{App:phi_n}. 

\subsection{The potential for the free parameters}
In this subsection, we will derive the potential for the free parameters.
By substituting 
\begin{align}
     f&=\int d^6y  \sqrt{g}\,G(x,y)\, e^{-\Phi+\frac{1}{2}\phi}
     \left(\frac{7}{9}\nabla^2 \phi+\frac{2}{9}b\right)
\end{align}
into the potential \eqref{makipotential2}, we get
\begin{align}
    V=\int d^6 x \sqrt{g} \left\{-\frac{17}{18}
    be^{\phi}
    -\frac{14}{9}e^{\phi}\nabla^2 \phi
    -e^{\Phi+\frac{1}{2}\phi}\left(\frac{7}{9}\nabla^2 \phi+\frac{2}{9}b\right)
    \int d^6y\sqrt{g} \,G(x,y)\, e^{-\Phi+\frac{1}{2}\phi} \left(\frac{7}{9}\nabla^2 \phi+\frac{2}{9}b\right)(y)
   \right\}.
    \label{makipotential3}
\end{align}
In the next step, we will substitute
\begin{align}
    \phi&=\sum_{i=0}^{\infty}\phi_i, \\
    \phi_n&=\left(-\frac{9}{2}\right)^n (-b)^{n+1} \sum_{i_1+i_2+\cdots+i_k=n+1}\frac{(-1)^k}{k}
     \Lambda_{i_1}\Lambda_{i_2}\cdots \Lambda_{i_k}, \\
     \Lambda_j(x)&=\int d^6y_1  \sqrt{g} \cdots \int d^6y_j  \sqrt{g}\,G(x,y_1)G(y_1,y_2)\cdots G(y_{j-1},y_j), 
\end{align}
and evaluate each term in the potential. Especially, we will approximate integrals in the potential by introducing cutoffs, in order to resolve difficulty of numerical analysis.  

\subsubsection{The first term}
We analyze the first term of Eq.\eqref{makipotential3}.
\begin{align}
  &-\frac{17}{18}\int d^6 x \sqrt{g}\, 
    be^{\phi}  \notag \\
    &=
   -\frac{17}{18} \int d^6 x \sqrt{g} \,
    b \sum_{n=0}^{\infty}\frac{1}{n!}\phi^n \notag \\
    &=-\frac{17}{18}\int d^6 x \sqrt{g} \,
    b \sum_{n=0}^{\infty}\frac{1}{n!}(\sum_{m=0}^{\infty}\phi_m)^n
    \notag \\
     &=-\frac{17}{18}\int d^6 x \sqrt{g} \,
    b \sum_{n=0}^{\infty}\frac{1}{n!}\left\{\sum_{m=0}^{\infty}\left(\left(-\frac{9}{2}\right)^m (-b)^{m+1} \sum_{i_1+\cdots+i_k=m+1}\frac{(-1)^k}{k}
     \Lambda_{i_1}\Lambda_{i_2}\cdots \Lambda_{i_k}\right)\right\}^n.
\end{align}
Here, we approximate the integral by introducing a cutoff, yielding the following expression
\begin{align}
    \int d^6 x \sqrt{g}
\Lambda_{j_1}(x)\Lambda_{j_2}(x)\cdots \Lambda_{j_k}(x)
    \sim \int d^6 x \sqrt{g} \,(-\bar{\Lambda})^{j_1+j_2+\cdots+j_k},
\end{align}
where $\bar{\Lambda}$ is a cutoff constant. A minus sign is placed in front of the cutoff $\bar{\Lambda}$ to ensure that $\bar{\Lambda}$ is positive, because the integral of the Green’s function is negative. For example, in the d-dimensional flat case, the Green's function is given by 
$G(x; x')= -\frac{1}{4} \pi^{-\frac{d}{2}}\Gamma(\frac{d}{2}-1) |x-x'|^{2-d}$ so that $\int_0^{r_0} d^d x G(x; 0)
=
-\frac{1}{2(d-2)}r_0^2$.
As a result, we get
\begin{align}
  -\frac{17}{18} \int d^6 x \sqrt{g} \,
    be^{\phi} &\sim
   -\frac{17}{18}\int d^6 x \sqrt{g} \,b e^{\bar{\phi}},
\end{align}
where
\begin{align}
    \bar{\phi}&=\sum_{n=0}^{\infty} \left(-\frac{9}{2}\right)^n (-b)^{n+1} (-\bar{\Lambda})^{n+1} \sum_{i_1+\cdots+i_k=n+1}\frac{(-1)^k}{k} \notag \\
    &=\frac{2}{9}\sum_{n=0}^{\infty}\frac{1}{n+1}\left(-\frac{9}{2}b\bar{\Lambda}\right)^{n+1}
    \\
    &=-\frac{2}{9}\log\left(1+\frac{9}{2}b\bar{\Lambda}\right).
\end{align}
According to the convergence condition, we have $\frac{1}{2}b\bar{\Lambda}<1$. In the second equality, we use the following relation:
\begin{align}
    \sum_{i_1+\cdots+i_k=n+1}\frac{(-1)^k}{k}=-\frac{1}{n+1},
    \label{co}
\end{align} 
whose detailed derivation is given in the Appendix \ref{App:co}.

\subsubsection{The second term}
Next, we analyze the second term of Eq.\eqref{makipotential3}.
\begin{align}
    &-\frac{14}{9}\int d^6 x \sqrt{g}\, e^{\phi}\nabla^2 \phi
    \notag \\
    &=
    -\frac{14}{9}\int d^6 x \sqrt{g}\, e^{\sum_{m=0}^{\infty}\phi_m}\nabla^2  \sum_{l=0}^{\infty}\phi_l \notag \\
    &= -\frac{14}{9}\int d^6 x \sqrt{g}\,
    \sum_{n=0}^{\infty}\frac{1}{n!}\left\{\sum_{m=0}^{\infty}\left(\left(-\frac{9}{2}\right)^m (-b)^{m+1} \sum_{i_1+\cdots+i_k=m+1}\frac{(-1)^k}{k}
     \Lambda_{i_1}\Lambda_{i_2}\cdots \Lambda_{i_k}\right)\right\}^n
\notag \\
   & \qquad \times \nabla^2
\sum_{l=0}^{\infty}\left(\left(-\frac{9}{2}\right)^l (-b)^{l+1} \sum_{j_1+\cdots+j_{k^{\prime}}=l+1}\frac{(-1)^{k^{\prime}}}{{k^{\prime}}}
     \Lambda_{j_1}\Lambda_{j_2}\cdots \Lambda_{j_{k^{\prime}}}\right) \notag \\
    &= -\frac{14}{9}\int d^6 x \sqrt{g}\,
    \sum_{n=0}^{\infty}\frac{1}{n!}\left\{\sum_{m=0}^{\infty}\left(\left(-\frac{9}{2}\right)^m (-b)^{m+1} \sum_{i_1+\cdots+i_k=m+1}\frac{(-1)^k}{k}
     \Lambda_{i_1}\Lambda_{i_2}\cdots \Lambda_{i_k}\right)\right\}^n
\notag \\
   & \qquad \times 
\sum_{l=0}^{\infty}\left(\left(-\frac{9}{2}\right)^l (-b)^{l+1} \sum_{j_1+\cdots+j_{k^{\prime}}=l+1}\frac{(-1)^{k^{\prime}}}{{k^{\prime}}}
     \Lambda_{j_1}\Lambda_{j_2}\cdots \Lambda_{j_{k^{\prime}-1}}\right).
\end{align}
In the third equality, we use the relation $\nabla^2 \Lambda_{i}=\Lambda_{i-1}$.
Here, we approximate the integral by
\begin{align}
\int d^6 x \sqrt{g} 
    \Lambda_{j_1}(x)\Lambda_{j_2}(x)\cdots \Lambda_{j_N}(x) \nabla^2 (\Lambda_{j_{N+1}}(x)\cdots \Lambda_{j_M}(x))
    \sim \int d^6 x \sqrt{g}(-\bar{\Lambda})^{j_1+j_2+\cdots+j_N+j_{N+1}+\cdots +j_{M}-1}.
\end{align}
 Then we get
\begin{align}
    -\frac{14}{9}\int d^6 x \sqrt{g}\, e^{\phi}\nabla^2 \phi
=-\frac{14}{9}\int d^6 x \sqrt{g}\,e^{\bar{\phi}} {\nabla^2\bar{\phi}},
\label{secondterm}
\end{align}
where 
\begin{align}
 {\nabla^2\bar{\phi}}=\frac{\bar{\phi}}{-\bar{\Lambda}}.
\end{align}

\subsubsection{The potential}
Similarly, the potential in \eqref{makipotential3} can be approximated as
\begin{align}
    V=\int d^6 x \sqrt{g}\left\{ -\frac{17}{18}
    be^{\bar{\phi}}
    +\frac{14}{9}e^{\bar{\phi}} \frac{\bar{\phi}}{\bar{\Lambda}}
    +{\bar{\Lambda}}\, e^{\bar{\phi}} \left(-\frac{7}{9}\frac{\bar{\phi}}{\bar{\Lambda}}+\frac{2}{9}b\right)^2\right\},
    \label{makipotential4}
\end{align}
where 
\begin{align}
    b&=\frac{\alpha^{\prime}}{2}e^{-\frac{1}{2}\Phi}{\rm{tr}}|F|^2, \notag \\
    \bar{\phi}&=-\frac{2}{9}\log\left(1+\frac{9}{2}b\bar{\Lambda}\right),
\end{align}
and ${\bar{\Lambda}}$ is a cutoff constant.

\subsection{Numerical analysis of the minimum of the potential for the parameters}
\begin{figure}[ht]
    \centering
   \includegraphics[width=0.6\textwidth]{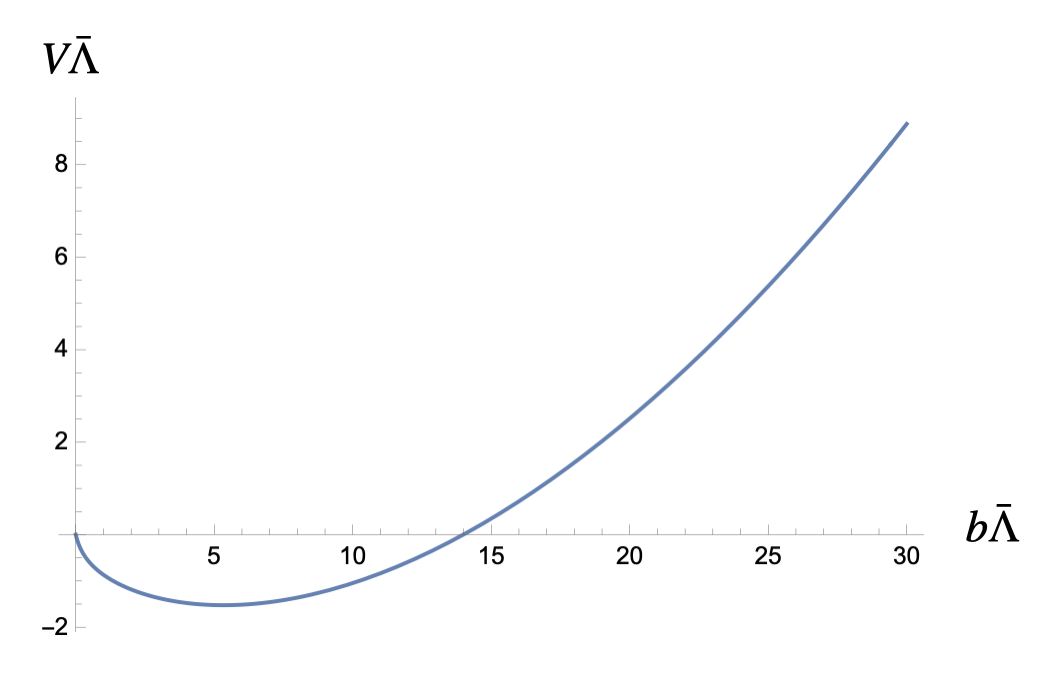}
    \caption{The minimum of the potential}
    \label{Fig;potential}
\end{figure}

In this subsection, we will perform a numerical analysis to investigate the minimum of the resulting potential:
\begin{align}
    \bar{\Lambda}V=\int d^6 x \sqrt{g}\,\left(1+\frac{9}{2}b\bar{\Lambda}\right)^{-\frac{2}{9}}\left\{
    -\frac{17}{18}
    b\bar{\Lambda}
    -\frac{28}{81}\log\left(1+\frac{9}{2}b\bar{\Lambda}\right)
    +\left(\frac{14}{81}\log\left(1+\frac{9}{2}b\bar{\Lambda}\right)+\frac{2}{9}b\bar{\Lambda}\right)^2\right\}.
\label{finalpotential}
\end{align}
This can be easily analyzed, and we obtain the result shown in Figure \ref{Fig;potential}. It can be shown that $b\bar{\Lambda}=5.32961$ corresponds to the minimum of the potential. 

\begin{figure}[ht]
    \centering
   \includegraphics[width=0.7\textwidth]{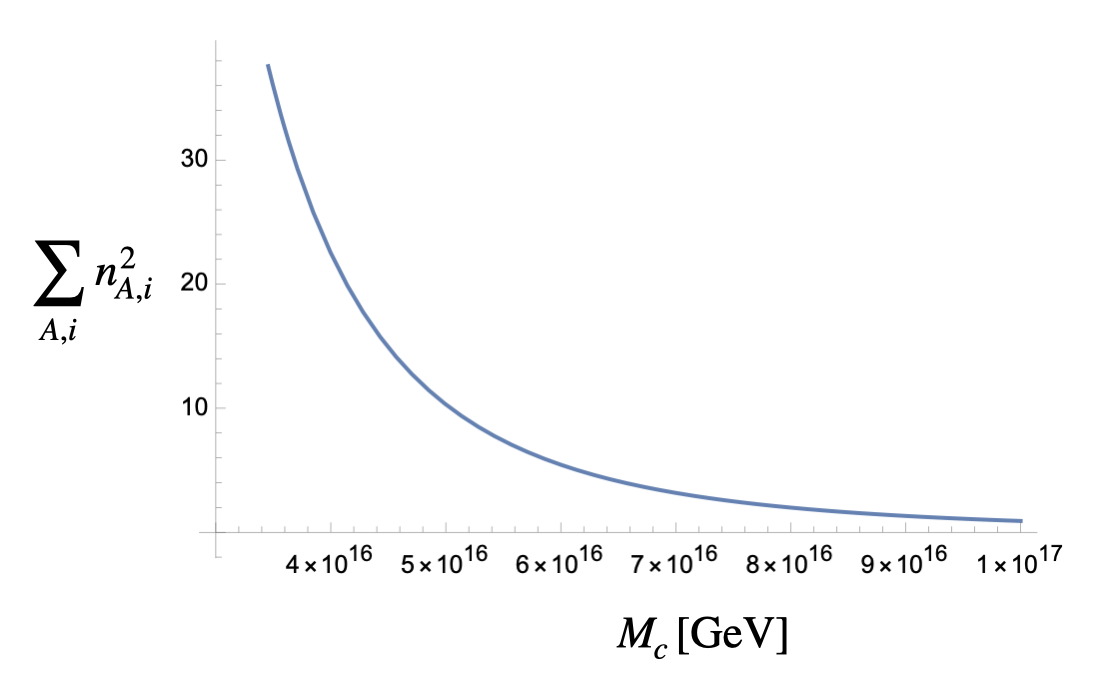}
    \caption{Constraint between the compactification scale and the flux quantization number}
    \label{Fig;constrain1}
\end{figure}

Here, we rewrite $b\bar{\Lambda}$ as
\begin{align}
    b\bar{\Lambda}&=\frac{V_2}{2M_s^2}e^{-\frac{1}{2}\Phi}{\rm{tr}}|F|^2
    ,
    \label{bLambda0}
\end{align}
where we identified $\bar{\Lambda}$ with a two-dimensional volume $V_2$.
The four-dimensional Planck mass $M_{\rm{pl}}$ arises from the dimensional reduction of the Einstein–Hilbert term on a six-dimensional internal manifold as
\begin{align}
    \frac{M_{\rm{pl}}^2}{16\pi}=\frac{V_2^3}{(2\pi)^7{\alpha^{\prime}}^4}e^{-2 \Phi }.
    \label{EH}
\end{align}
Thus, we obtain
\begin{align}
    b\bar{\Lambda}
    &=\frac{V_2}{2M_s^2}\frac{\sqrt{M_{\rm{pl}}}}{\left(\frac{16\pi}{(2\pi)^7}\right)^{\frac{1}{4}}M_s^2 V_2^{\frac{3}{4}}}
    {\rm{tr}}|F|^2.
    \label{bLambda}
\end{align}
On the other hand, since the four-dimensional gauge coupling is obtained through dimensional reduction from the ten-dimensional gauge field strength
\begin{align}
    \alpha^{-1}=\frac{4\pi}{g_4^2}=\frac{4\pi V_2^3}{4(2\pi)^7{\alpha^{\prime}}^3}e^{-2 \Phi }=
    \frac{M_{\rm{pl}}^2}{16M_{\rm{s}}^2},
\end{align}
where we used the relation \eqref{EH} and the string scale $M_{\rm{s}}^2=1/\alpha^{\prime}$.
Here, $\alpha$ is a unified fine structure constant, i.e. $\alpha \simeq 1/24$ in a GUT-like setting. Therefore, by substituting the Planck scale $M_{\rm{pl}}=1.2\times 10^{19} \,{\rm{GeV}}$, the string scale $M_{\rm{s}}$ is fixed at $6.1\times 10^{17} \rm{GeV}$.
From Eq.\eqref{bLambda}, since the free parameters contained in $b\bar{\Lambda}$ are ${\rm{tr}}|F|^2$ and $V_2$, the minimum of the potential leads to a nontrivial following relation between ${\rm{tr}}|F|^2$ and $V_2$:
\begin{align}
    \frac{\sqrt{M_{\rm{pl}}}}{2\left(\frac{16\pi}{(2\pi)^7}\right)^{\frac{1}{4}}M_s^4 }V_2^{\frac{1}{4}}
    {\rm{tr}}|F|^2=5.32961.
\end{align}
From Eq.\eqref{efv}, because the gauge field strength is quantized, ${\rm{tr}}|F|^2$ is expressed as
\begin{align}
    {\rm{tr}}|F|^2&=\sum_{A=1}^{11}\sum_{i=1}^{3}
    4\left(\frac{2\pi n_{Ai}}{V_2}\right)^2 \\
    &=\sum_{A=1}^{11}\sum_{i=1}^{3}
    4(2\pi n_{Ai}M_c^2)^2.
\end{align}
Here, $V_2\sim 1/M_c^2$, and $M_c\,\rm{[GeV]}$ denotes the compactification scale.
Expressing $b\bar{\Lambda}=5.32961$ in terms of the flux quantization number and the compactification scale
\begin{align}
   \sum_{A=1}^{11}\sum_{i=1}^{3} n_{Ai}^2=\frac{4.5\times 10^{61} {\rm{[GeV^{\frac{7}{2}}]}}}{16\pi^2}M_c^{-\frac{7}{2}}.
\end{align}
The result is presented in Fig.\ref{Fig;constrain1}.
Note that since the flux quantization number is an integer, the vertical axis of Fig.\ref{Fig;constrain1} only takes possible integer values. From Eq.\eqref{engen}, the number of generations is calculated as
\begin{align}
N_{\text{gen}}=&\left|\prod_{i=1}^3\sum_{A=1}^3 n_{Ai} -\sum_{A=1}^3 \prod_{i=1}^3n_{Ai}\right|. \label{engen2}
\end{align}
By systematically examining the cases starting from 
$\sum_{A=1}^{11}\sum_{i=1}^{3} n_{Ai}^2=1$, we find through brute-force search of the combinations that yield three generations from Eq.\eqref{engen2} that the minimal value is 
$\sum_{A=1}^{11}\sum_{i=1}^{3} n_{Ai}^2=5$.
Specifically, it takes the following form:
\begin{align}
    n_{Ai}=
    \begin{pmatrix}
1 & 0 & 0 \\
1 & 1 & 0 \\
1 & 0 & 1 \\
\vec{0} & \vec{0} &\vec{0} 
\end{pmatrix}.
\end{align}
Therefore, for a model that accommodates three generations, at least $\sum_{A=1}^{11}\sum_{i=1}^{3} n_{Ai}^2=5$ is required.
For 
$\sum_{A=1}^{11}\sum_{i=1}^{3} n_{Ai}^2=5$, the compactification scale is found to be
$6.1 \times 10^{16}\,\rm{GeV}$, which leads to the following constraint
\begin{align}
    M_{c} \le 6.1 \times 10^{16} \,{\rm{GeV}} < M_s=6.1\times 10^{17}\,\rm{GeV}.
    \label{comconstrain}
\end{align}



\section{Conclusion and discussion}
\label{sec:conclusion}

In this paper, we have analyzed minima of the potential for string backgrounds obtained from string geometry theory \cite{Nagasaki:2023fnz, Nagasaki:2025tmi, Kudo}. Especially, we have studied the region specified by a heterotic model assuming a constant dilaton and the presence of field strength only \cite{Tsuyuki:2021xqu,Takeuchi:2023egl}.

In section \ref{sec:heterocom}, we have briefly reviewed the heterotic sector of string geometry theory. Especially, we have given the explicit potential for heterotic string backgrounds in the Einstein frame with the warped compactification in a particle limit $\bm X_{G}^{(\mu\bar\sigma\bar\theta)}  \to x^{\mu}$. One can use this potential to determine free parameters of heterotic string phenomenological models.

In section \ref{sec:setup}, we have reviewed the model representing a non-supersymmetric compactifications in heterotic supergravity. As discussed in \cite{Takeuchi:2023egl}, the number of generations is constrained by the equation of motion and the anomaly cancellation mechanism.
In the case where the Euler characteristics of the three internal submanifolds are $(\chi_1,\chi_2,\chi_3)=(2,-2,-2)$, the number of generations is restricted to three or fewer. However, as the genus increases, the allowed values of the flux also become larger, and thus the constraint on the number of generations becomes weaker.

In Section \ref{sec:calculationpotential}, we have investigated which solution should be selected as a grand state among the above compactifications that satisfy the equations of motion and the anomaly cancellation condition, by examining the minima of the potential in that region.
We have obtained the minimum of the potential for the parameters shown in Fig.\ref{Fig;potential}, which is obtained by substituting the model to the potential for string backgrounds. 
Consequently, the values of the compactification scale and the flux quanta are constrained as illustrated in Figure \ref{Fig;constrain1}.
For a model that accommodates three generations, the compactification scale should satisfy $M_c < M_s$. Namely, compactifications can be described by the supergravity and its corrections.

In this analysis, the values of the dilaton and the field strength, treated as free parameters, could not be fixed independently. This limitation arises from our assumption of a constant dilaton: when $f$ is rewritten in terms of $e^{-\Phi}\tilde{f}$ in Eq.\eqref{makipotential}, \eqref{phieq} and \eqref{feq}, the dilaton and the field strength appear only in the combintaion (\ref{combination}). 
If the assumption of a constant dilaton is relaxed, contributions from the derivatives of the dilaton in Eq.\eqref{makipotential}, \eqref{phieq} and \eqref{feq} would appear, allowing the dilaton and the field strength to be treated independently.
Besides, in case that $\tilde{H} \neq 0$, not only ${\rm{tr}}|F|^2$, but also the components of the fluxes $F$ would be determined.


In similar analyses to this paper's, one can determine parameters in other string phenomenological models, identify the ground states in the models, and obtain their values of the potential energy. By comparing them, one can find a model near the true vacuum in a sense of energy. In these analyses, naturalness is not necessary, and thus the parameters in the models can be extended maximally, because there is a fundamental principle, the minimum of the potential. An abandoned model might include the true vacuum. By comparing various models, the true vacuum in string theory will be discovered.

\section*{Acknowledgments}
The authors thank the Yukawa Institute for Theoretical Physics at Kyoto University, where this work was initiated during the YITP-W-22-09 on ``Strings and Fields 2022.'' This work was supported by YAMAGUCHI UNIVERSITY FUND(M.T.) and the Sumitomo Foundation under Grant for Basic Science Research (Grant No. 2502479)(M.T.).

\appendix

\section{Detailed calculations to solve the condition for $\phi$}
\label{App:phi_n}

In this appendix, we demonstrate how $\phi_n$ in \eqref{phi_n} is derived.
Eq.\eqref{phi_i} is given by
\begin{align}
     \phi_i=-\frac{1}{4} \sum_{k+l+1=i} \phi_k  \phi_l +\frac{1}{2} \int d^6y \, G(x,y) \sum_{k+l+1=i} \phi_k (y)\nabla^2 \phi_l(y).
\end{align}
By substituting $i=1$, we obtain
\begin{align}
     \phi_1&=-\frac{1}{4} ( \phi_0 )^2 +\frac{1}{2} \int d^6y \, G(x,y)  \phi_0 (y)\nabla^2 \phi_0(y) \notag \\
     &=-\frac{1}{4}\left(b\int d^6y \,G(x,y)\right)^2
     +\frac{1}{2}b^2 \int d^6y\, d^6z\, G(x,y)G(y,z)
     \notag \\
     &=-\frac{1}{2}b^2\left(\frac{1}{2}\Lambda_1^2(x)-\Lambda_2(x)\right).
     \label{phi1}
\end{align}
In the first equality, we used the relation $\nabla^2 \phi_0=b$. The second equality results from substituting Eq.\eqref{phi0}, and the third equality results from replacing it with $\Lambda_j(x)$.
Next, we consider the case of $i=2$. By substituting $i=2$, we obtain
\begin{align}
     \phi_2&=-\frac{1}{2}  \phi_0 (y) \phi_1(y) +\frac{1}{2} \int d^6y \, G(x,y)  \left(\phi_0 (y)\nabla^2 \phi_1(y)+\phi_1 (y)\nabla^2 \phi_0(y)\right)
     \notag \\
     &=-\frac{1}{2}  \phi_0 (y) \phi_1(y) +\frac{1}{2} \int d^6y \, G(x,y) \left\{\phi_0 (y)\left(-\frac{1}{2}b^2 (\nabla \Lambda_1)^2\right)+b \phi_1 (y)\right\}
     \notag \\
     &=\left(-\frac{1}{2}\right)^2 b^3 \left(\frac{1}{2}\Lambda_1^3(x)-\Lambda_1(x)\Lambda_2(x)\right) 
     \notag \\
     & \qquad -\left(\frac{1}{2}\right)^2b^3 \int d^6y \, G(x,y) \left\{ \Lambda_1(y)(\nabla \Lambda_1(y))^2+\left(\frac{1}{2}\Lambda_1^2(y)-\Lambda_2(y)\right)\right\},
     \label{phi2_1}
\end{align}
where we used the relation $\nabla^2 \phi_1=-\frac{1}{2} b^2 (\nabla \Lambda_1)^2$ in the second equality, and the third equality results from Eqs.\eqref{phi0} and \eqref{phi1}.
We proceed by further transforming the second term of \eqref{phi2_1}.

\begin{align}
    &\int d^6y \, G(x,y)  \Lambda_1(y)(\nabla \Lambda_1(y))^2
    \notag \\
    &=-\int d^6y \, \nabla_y G(x,y) \Lambda_1^2(y)\nabla \Lambda_1(y)
    -\int d^6y \,  G(x,y) \Lambda_1(y)(\nabla \Lambda_1(y))^2
    -\int d^6y \,  G(x,y) \Lambda_1^2(y)\nabla^2 \Lambda_1(y)
    \notag \\
    &=\frac{1}{2}\left(-\int d^6y \, \nabla_y G(x,y) \Lambda_1^2(y)\nabla \Lambda_1(y)
    -\int d^6y \,  G(x,y) \Lambda_1^2(y)\nabla^2 \Lambda_1(y)
    \right)
    \notag \\
    &=\frac{1}{2}\int d^6y \, \nabla_y^2 G(x,y) \Lambda_1^3(y)+
    \int d^6y \, \nabla_y G(x,y) \Lambda_1^2(y)\nabla \Lambda_1(y)
    -\frac{1}{2}\int d^6y \,  G(x,y) \Lambda_1^2(y)\nabla^2 \Lambda_1(y)
    \notag \\
    &=\frac{1}{6}\int d^6y \, \nabla_y^2 G(x,y) \Lambda_1^3(y)
    -\frac{1}{2}\int d^6y \,  G(x,y) \Lambda_1^2(y)\nabla^2 \Lambda_1(y)
    \notag \\
    &=\frac{1}{6}\Lambda_1^3(x)-\frac{1}{2}\int d^6y \,  G(x,y) \Lambda_1^2(y).
    \label{A4}
\end{align}
In the last equality, we used the relation $\nabla^2 \Lambda_{1}=1$.
By substituting Eq.\eqref{A4} into Eq.\eqref{phi2_1}, we get
\begin{align}
     \phi_2&=-\left(-\frac{1}{2}\right)^2 b^3\left(-\Lambda_3(x)+\Lambda_2(x)\Lambda_1(x)-\frac{1}{3}\Lambda_1^3(x)\right).
     \label{phi_2}
\end{align}
Proceeding with the calculation in a similar manner, we obtain 
\begin{align}
\phi_n&=\left(-\frac{1}{2}\right)^n (-b)^{n+1} \sum_{i_1+i_2+\cdots+i_k=n+1}\frac{(-1)^k}{k}
     \Lambda_{i_1}\Lambda_{i_2}\cdots \Lambda_{i_k}.
     \label{Aphi_n}
\end{align}

\section{Detailed calculations of Eq.\eqref{co}}
\label{App:co}

In this appendix, we display the relation 
\begin{align}
    \alpha_n\equiv \sum_{i_1+\cdots+i_k=n+1}\frac{(-1)^k}{k}=-\frac{1}{n+1},
    \label{Aco}
\end{align} 
explicitly for the case $n=0, 1, \cdots, 5$. In the case of $n=0$, since the only possible value for the sum is $i_1=1$, we get
\begin{align}
    \alpha_0=-1,
\end{align}
and then, for the case $n=1$, the possible values for the sum are $i_1=2$ and $(i_1,i_2)=(1,1)$, so
\begin{align}
    \alpha_1=-1+\frac{1}{2}=-\frac{1}{2}.
\end{align}
Similarly, the possible combinations of $i_k$ are summarized in Table \ref{App:T1}.

\begin{table}[htbp]
  \centering
\begin{tabular}{|c|c|c|c|c|c|}
  \hline
   & $k=1$ & $k=2$ & $k=3$& $k=4$& $k=5$ \\
  $n$ & $i_1$ & $(i_1,i_2)$ & $(i_1,i_2,i_3)$& $(i_1,i_2,i_3,i_4)$& $(i_1,i_2,i_3,i_4,i_5)$ \\
  \hline
  $0$ & $1$ & & &&\\
  \hline
  $1$ & $2$ & $(1,1)$ &&& \\
  \hline
  $2$ & $3$ & $(1,2),(2,1)$&$(1,1,1)$ &&\\
  \hline
  $3$ & $4$ & $(1,3),(3,1),(2,2)$ &$(1,1,2),(1,2,1),(2,1,1)$&$(1,1,1,1)$&\\
  \hline
  $4$ & $5$ & $(1,4),(4,1),$ &$(1,1,3),(1,3,1),(3,1,1),$ &$(2,1,1,1),(1,2,1,1),$&$(1,1,1,1,1)$\\
  &&$(2,3),(3,2)$&$(1,2,2),(2,1,2),(2,2,1)$&$(1,1,2,1),(1,1,1,2)$& \\
  \hline
\end{tabular}
\caption{The possible combinations of $i_k$}
\label{App:T1}
\end{table}

From Table \ref{App:T1}, we get
\begin{align}
    \alpha_2&=(-1)+2\frac{(-1)^2}{2}+\frac{(-1)^3}{3}=-\frac{1}{3},
    \\
    \alpha_3&=(-1)+3\frac{(-1)^2}{2}+3\frac{(-1)^3}{3}+\frac{(-1)^4}{4}=-\frac{1}{4},
    \\
    \alpha_4&=(-1)+4\frac{(-1)^2}{2}+6\frac{(-1)^3}{3}+4\frac{(-1)^4}{4}+\frac{(-1)^5}{5}=-\frac{1}{5},
    \\
    \alpha_5&=(-1)+5\frac{(-1)^2}{2}+10\frac{(-1)^3}{3}+10\frac{(-1)^4}{4}+5\frac{(-1)^5}{5}+\frac{(-1)^6}{6}=-\frac{1}{6}.
\end{align}

\bibliographystyle{utphys}
\bibliography{ref}

\end{document}